\documentclass[aps,
twocolumn, 
showpacs,
prx,
superscriptaddress,
tightenlines,
10pt,
longbibliography] 
{revtex4-2}
\usepackage{graphicx}
\usepackage{xcolor}
\usepackage{braket}
\usepackage{hyperref}
\usepackage{comment}

\usepackage{siunitx}
\makeatletter
\newcommand*{\defeq}{\mathrel{\rlap{%
                     \raisebox{0.3ex}{$\m@th\cdot$}}%
                     \raisebox{-0.3ex}{$\m@th\cdot$}}%
                     =}
\makeatother

\UseRawInputEncoding

\begin{document}

\title{Ultra-low noise quantum memory for quasi-deterministic single photons generated by Rydberg collective atomic excitations}
\author{L. Heller}
\thanks{These two authors contributed equally}
\affiliation{ICFO-Institut de Ciencies Fotoniques, The Barcelona Institute of Science and Technology}
\author{J. Lowinski}
\thanks{These two authors contributed equally}
\affiliation{ICFO-Institut de Ciencies Fotoniques, The Barcelona Institute of Science and Technology}
\author{K. Theophilo}
\affiliation{ICFO-Institut de Ciencies Fotoniques, The Barcelona Institute of Science and Technology}
\author{A. Padr\'on-Brito}
\affiliation{ICFO-Institut de Ciencies Fotoniques, The Barcelona Institute of Science and Technology}
\author{H. de Riedmatten}
\affiliation{ICFO-Institut de Ciencies Fotoniques, The Barcelona Institute of Science and Technology}
\affiliation{ICREA-Instituci\'o Catalana de Recerca i Estudis Avan\c cats, 08015 Barcelona, Spain}

\begin{abstract}
We demonstrate the storage and retrieval of an on-demand single photon generated by a collective Rydberg excitation in an ultra-low noise Raman quantum memory located in a different cold atomic ensemble. We generate single photons on demand by exciting a cold cloud of Rubidium atoms off resonantly to a Rydberg state, with a generation probability up to 15 $\%$ per trial. We then show that the single photons can be stored and retrieved with an efficiency of 21 $\%$ and a noise floor of $p_n= 2.3(3) \times 10^{-4}$ per trial in the Raman quantum memory. This leads to a signal-to-noise ratio ranging from 11 to 26 for the retrieved single photon depending on the input photon generation probability, which allows us to observe significant antibunching. We also evaluate the performances of the Raman memory as built-in unbalanced temporal beam splitter, tunable by varying the write-in control pulse intensity. In addition, we demonstrate that the Raman memory can be used to control the single-photon waveshape. These results are a step forward in the implementation of efficient quantum-repeater links using single-photon sources. 
\end{abstract}

\maketitle

\section{Introduction}
\label{sec:introduction}
The development of quantum networks \cite{Kimble2008} is one of the most active branches of quantum technologies.
A fundamental step towards long distance quantum communications would be the realization of efficient quantum repeaters \cite{Briegel1998, Duan2001, Sangouard2011} that would allow the distribution of entanglement over distances longer than what is achievable with direct photon transmission. 
Near term quantum repeaters are based on heralded entanglement between remote quantum memories that form an elementary repeater link and on entanglement swapping between the elementary links to extend the entanglement distance.
Various platforms have been considered as quantum memories for quantum repeaters, including single atoms \cite{Ritter2012, Langenfeld2021} or ions \cite{Stute2012, Hucul2015, Krutyanskiy2019, Stephenson2020}, cold atomic clouds \cite{Chaneliere2005, Chou2005, Chou2007, Yuan2008, Cao2020, Heller2020, Yu2020} and solid-state systems such as color centers in diamond \cite{Humphreys2018, Bhaskar2020} or rare-earth doped solids \cite{Riedmatten2008, Hedges2010, Saglamyurek2011, Clausen2011, Lago-Rivera2021, Liu2021}.
Ensemble-based approaches are promising for reaching high-rate quantum repeaters because they enable large multiplexing, which increases the rate of entanglement in elementary links proportionally to the number of stored modes \cite{Simon2007, Collins2007, Lan2009, Pu2017, Chang2019, Heller2020, Lago-Rivera2021}. 

To date, most of the early demonstrations of quantum repeater links with ensemble-based quantum memories are based on probabilistic light-matter entanglement sources, e.g. based on emissive quantum memories using spontaneous Raman scattering in atomic clouds \cite{Chou2005, Chou2007, Felinto2006, Yuan2008, Yu2020}, following the DLCZ proposal, or by using read-write quantum memories combined with spontaneous parametric down-conversion sources \cite{Lago-Rivera2021, Liu2021}.
However, this type of probabilistic sources leads to limitations due to a trade-off between excitation probability and fidelity of the generated state.
To keep the errors due to the generation of multiple pairs low, and therefore the fidelity high, the generation probability must remain low.
This trade-off leads to low success probability per trial (especially for multiple-link repeaters), which limits the overall high-fidelity entanglement rate \cite{Sangouard2011}.

A quantum repeater architecture based on the use of deterministic single photons and absorptive ensemble-based quantum memories was proposed to overcome this limitation \cite{Sangouard2007}.
In this scheme, each node consists of a deterministic single-photon source and a quantum memory.
The single photon is sent on a beam splitter (BS) and one output of the BS is directed towards the quantum memory while the other output is converted to telecom wavelength and sent to a central station where it is mixed with the photonic mode from another distant quantum node.
It has been shown that heralded single photons (generated from probabilistic sources) can be stored in quantum memories \cite{Seri2017, Ding2015b} with up to \SI{87}{\percent} storage and retrieval efficiency \cite{Wang2019, Cao2020}.
Hence, the main challenge of this scheme compared to schemes using probabilistic sources is to generate memory-compatible indistinguishable single photons on demand with high efficiency.
In addition, the quantum memory should feature very low noise in order not to degrade the single photon properties. 

Several approaches have been demonstrated to generate on-demand single photons using single emitters such as quantum dots, single molecules and color centers in diamond.
However, most of these photons are not resonant with quantum memories and have a bandwidth much larger than the one of long-lived quantum memories.
While progress has been made recently to interface photons from quantum dots and molecules to atomic vapors or rare-earth doped solids \cite{Akopian2011, Siyushev2014, Tang2015, Kroh2019}, so far high efficiency and long-lived storage of these photons has not been demonstrated.  
Single trapped atoms can be used to generate directly resonant and memory-compatible photons that have been interfaced with a BEC quantum memory \cite{Lettner2011}, however, the efficient photon generation in a single mode requires placing the atom in a high-finesse cavity, which represents an experimentally complex task.
In recent years, several experiments have shown that ensembles of Rydberg atoms could serve as a source of on-demand narrow-band \cite{Dudin2012, Maxwell2013, Li2016, Li2019} indistinguishable single photons \cite{Li2013, OrnelasHuerta2020, Padron-Brito2021}.
This approach has the advantage that no high-finesse cavity is required, due to the collective nature of the single photon generation. 

In this paper, we demonstrate the storage and retrieval of an on-demand single photon generated by a collective Rydberg excitation on a ultra-low noise Raman quantum memory located in a different cold atomic ensemble.
We show that the single photons can be stored and retrieved with a signal-to-noise ratio (SNR) up to 26, preserving strong antibunching.
We also evaluate the performance of the built-in temporal beam splitter offered by the Raman memory.  
In addition, we demonstrate that the Raman memory can be used to control the single photon waveshape.
These results show that single photons generated on demand by Rydberg atoms can be stored in an atomic quantum memory, which is an important step towards the implementation of efficient quantum-repeater links using single-photon sources.

\section{Experimental Setup}
\label{sec:experimental_setup}

\begin{figure*}
	\includegraphics[width=0.99\linewidth]{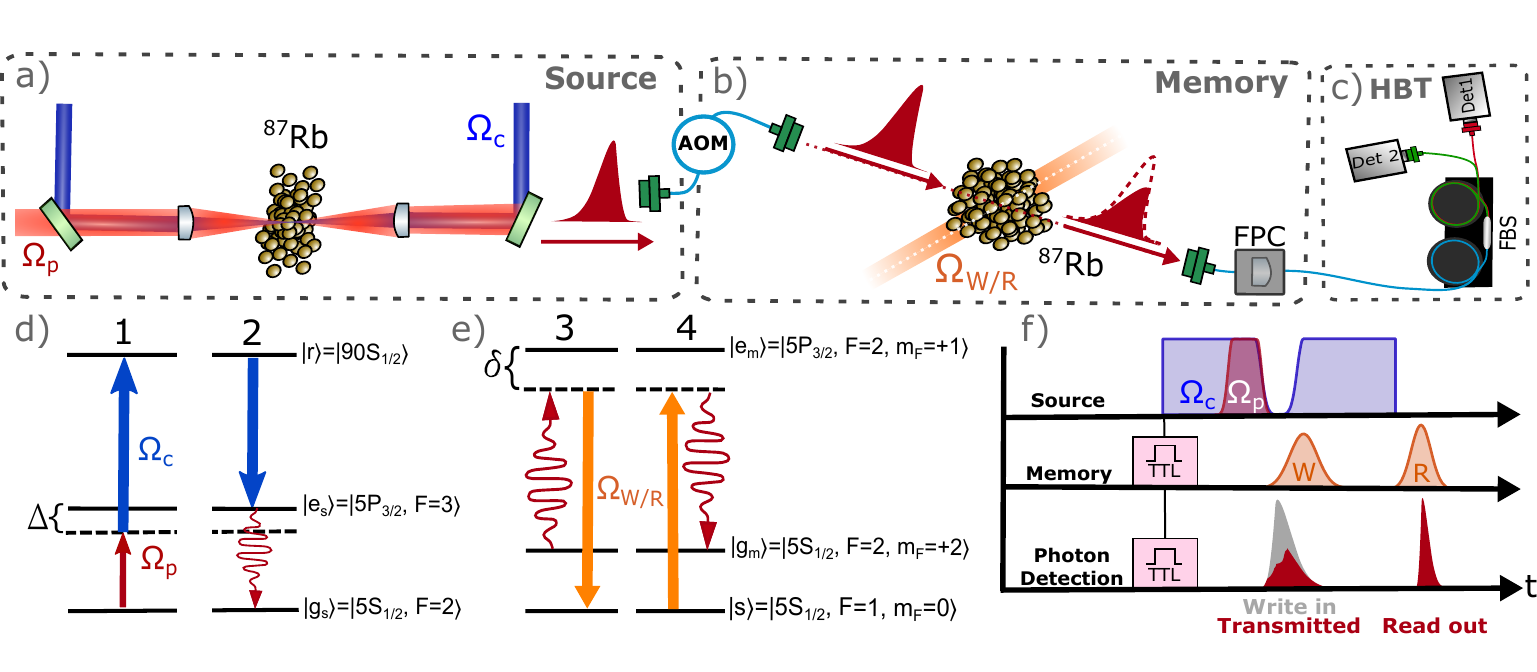}
	\caption{A scheme of the experimental setup, the relevant atomic levels and the experimental sequence.
    (a) The source.
    The probe ($\Omega_p$) and the counter-propagating coupling beam ($\Omega_c$) are tightly focused in a cold cloud of Rubidium atoms to generate the input photon.
    (b) The memory.
    A write-in control beam pulse ($\Omega_W$) maps the incoming photon to an atomic excitation in another cold cloud of Rubidium atoms. The excitation is retrieved with a read-out control beam pulse ($\Omega_R$) and filtered with a Fabry-Perot cavity (FPC).
    (c) The retrieved photon is split in a fiber-based beam-splitter (FBS) and detected with two superconducting nanowire single-photon detectors SNSPD 1 and 2 performing, effectively, an HBT measurement.
    The relevant atomic levels for the photon generation (d) and for the photon storage (e) are also shown. A two-photon excitation ($\Omega_p$ and $\Omega_c$, with $\Delta = \SI{-40}{\mega\hertz}$) creates a Rydberg spin wave in $\ket{r}$ (1) which is later mapped to the first excited state $\ket{e_s}$ and decays emitting a photon (2). 
    The emitted photon is mapped with $\Omega_W$ to a ground-state spin wave in $\ket{s}$ (3) and later retrieved with $\Omega_R$ (4).
    (f) The pulse sequence.
    The whole experiment is synchronized with TTLs sent by the source at the beginning of each generation trial.}
	\label{fig:ramanexpsk}
\end{figure*}

Our experimental setup comprises of two ensembles of cold $^{87}$Rb atoms situated in the same laboratory connected via \SI{12}{\meter} of optical fiber cable.
One of them is used to generate single photons in a quasi-deterministic way by exploiting the strong dipole-dipole interaction between Rydberg states (the source).
Another is used to store and on-demand retrieve the generated photons in an atomic Raman memory.

In the first step of the generation protocol we excite the ensemble from its ground state $\ket{g_{s}} = \ket{5S_{1/2}, F=2}$ to a Rydberg state $\ket{r} = \ket{90S_{1/2}}$, see \autoref{fig:ramanexpsk}(d), via a two-photon excitation.
We send a weak coherent probe pulse $\Omega_p$ and a strong counter-propagating coupling pulse $\Omega_c \approx \SI{6}{\mega\hertz}$, see \autoref{fig:ramanexpsk}(a).
The probe light at a wavelength of \SI{780}{\nano\meter} is red-detuned by \SI{-40}{\mega\hertz} from the transition to the excited state $\ket{e_{s}} = \ket{5P_{3/2}, F=3}$.
The coupling light is tuned such that the two-photon transition is resonant with the transition $\ket{g_s} \rightarrow \ket{r}$.

The number of generated Rydberg excitations is strongly limited due to the dipole blockade \cite{Lukin2001}.
The blockade is a result of the strong dipole-dipole interaction between Rydberg states, which prevents a simultaneous excitation of two Rydberg atoms, if they are closer than a distance called the blockade radius.
Then, if the interaction region is smaller than the volume given by the blockade radius, only one atomic excitation will be created in state $\ket{r}$ - this is called the fully blockaded regime.
The Rydberg excitation is shared between all the atoms in the blockade region, forming a collective quantum superposition, termed Rydberg spin wave.

With a delay of \SI{1}{\micro\second}, a second coupling pulse is sent resonantly to the $\ket{r} \rightarrow \ket{e_s}$ transition, mapping the Rydberg spin wave onto the excited state $\ket{e_s}$ and triggering the collective emission of a single photon at \SI{780}{\nano\meter}.
The photon is emitted in the input mode and in forward direction thanks to collective atomic interference.
It is then separated from the coupling light by a dichroic mirror and a band-pass filter, before being collected into a polarization-maintaining single-mode fiber.
An electronic trigger is sent to the memory to signal each photon generation attempt.

The generated photon is guided to the second atomic ensemble, the memory.
The frequency of the photon is, however, not compatible with the transitions used in the memory, so it is shifted by \SI{-320}{\mega\hertz} with an acousto-optic modulator (AOM).
As a result, the photon is now red-detuned with respect to the $\ket{g_m} \rightarrow \ket{e_m}$ transition.

The Raman memory relies on coherent, adiabatic absorption of the incoming single photon \cite{Nunn2007}. 
A storage attempt starts with sending a control write-in pulse $\Omega_W$ coupling states $\ket{s} = \ket{5S_{1/2}, F=1, m_F=0}$ and $\ket{e_m} = \ket{5P_{3/2}, F=2, m_F=+1}$ off-resonantly by $\delta = \SI{-52}{\mega\hertz}$, see \autoref{fig:ramanexpsk}(e).
Since the write-in pulse is in two-photon resonance with the input photon, the incoming photon field is transferred to a collective atomic spin excitation on $\ket{s}$.
Careful tuning of the control write-in pulse shape, power and timing with respect to the input photon is required to optimize the writing efficiency into the memory.

To retrieve the stored excitation, after a programmable delay, we send a read-out pulse $\Omega_R$.
The read-out pulse is in the same spatial mode as the write-in pulse with the same frequency detuning $\delta$.
Owing to the collective atomic interference, the photon is emitted in the input mode in forward direction and collected into a single-mode fiber.
The bandwidth and the shape of the output photon are governed by the temporal profile and the power of the read-out pulse and can be tuned arbitrarily (see \autoref{sec:photon_storage}).

The collected photons are guided to the detection setup.
Depending on the measurement it is either a superconducting nanowire single-photon detector (SNSPD) or a Hanbury Brown-Twiss (HBT) setup comprised of a fiber-based beam splitter and two SNSPDs, see \autoref{fig:ramanexpsk}(c).
We use HBT setup to measure photons autocorrelation.

For the source to produce single photons it is necessary to be in the fully blockaded regime or close to it.
Moreover, the coherence time of the Rydberg transition should be much longer than the excitation pulse duration to keep the photon generation probability high.
In our case, this means that the ensemble needs to be cold to limit the motional dephasing.
Both are achieved in the preparation stage when we first load the atoms into a magneto-optical trap (MOT), later compress them and subsequently apply \SI{7}{\milli\second} of polarisation-gradient cooling.
Finally the ensemble is prepared by optical pumping to its initial ground state $\ket{g_{s}} = \ket{5S_{1/2}, F=2}$.
A one-dimensional dipole trap is kept on during the whole process (with a beam waist of \SI{34}{\micro\meter} at an angle of \ang{22} with respect to the probe beam and a depth of \SI{0.3}{\milli\kelvin}).
The whole process results in a cloud with OD of 6 and a temperature of \SI{40}{\micro\kelvin}.
Thanks to the dipole trap, the effective interaction region, given by the overlap between the probe beam and the atomic ensemble, is still larger but comparable to the $\sim \SI{13}{\micro\meter}$ of the blockade radius.
The ensemble can be interrogated for \SI{200}{\milli\second}, limited by the population lifetime of the dipole trap (\SI{400}{\milli\second}), before another MOT reloading cycle has to be performed.
During its interrogation time, the source attempts to generate a single photon every \SI{4}{\micro\second} with generation probability $p_{\mathrm{gen}} = \SIrange{5}{15}{\percent}$.
The photon generation and the characterization of their indistinguishability is described in more detail in \cite{Padron-Brito2021}.

For the memory, the OD of the ensemble and the cloud temperature are the main parameters governing the storage and retrieval efficiency and the storage time.
To achieve a dense and cold ensemble the atoms are first loaded into a MOT for \SI{10}{\milli\second} followed by \SI{1.5}{\milli\second} of polarisation-gradient cooling.
Later the memory is optically pumped to its initial ground state $\ket{g_{m}} = \ket{5S_{1/2}, F=2, m_F=+2}$, in the presence of a homogeneous magnetic bias field oriented along the photon mode.
Optical pumping is helpful to avoid beating between spin waves at different Zeeman sublevels.
The whole process provides us a cloud with OD of 5 and a temperature of \SI{30}{\micro\kelvin}.
OD starts dropping after \SI{1.2}{\milli\second} of interrogation time and the trapping cycle has to be repeated.

One of the main challenges of this study were long integration times which required good stability of both setups.
This results mainly from two technical limitations.
The first one are very different trapping cycles of the source and the memory making the overall duty cycle very low.
The resulting repetition rate of the whole experiment is \SI{5}{\kilo\hertz}.
The second one is the passive loss in the photon transmission which affects quadratically the coincidence probability in the HBT experiment.
The total transmission from the output of the source to the detection setup, in the absence of atoms in the memory, is \SI{10(1)}{\percent}, limited by the fiber coupling after the source (0.4), the frequency-shifter AOM setup (0.62), the fiber coupling after the memory (0.83), the frequency-filtering cavity setup (0.65) and miscellaneous optical and polarisation-dependent losses (0.75).
The transmission from the output of the source to the input of the quantum memory is 22 $\%$.
The SNSPDs have quantum efficiency $\sim \SI{85}{\percent}$ and \SI{3}{\hertz} of dark counts.

The limiting factor for the quality of the single photon retrieved from the quantum memory is the introduced technical noise, which affects its SNR.
The main source of noise is the leakage of the memory control pulses which couple to the photon mode.
An angle of 3 degrees between the photon mode and the coupling beam minimizes the spatial overlap and noise introduced by directional, forward scattering.
The noise is further removed with a home-built narrow-band Fabry-Perot filter cavity of 43.4 dB suppression (at the control pulse frequency).
The remaining noise is composed of light leaking through the filter, inelastically scattered control light at the photon frequency and the detectors' dark counts.

\section{Results}
\label{sec:results}
In this section we study the single photon properties of the source photons which, further on, are used as the memory input photon. 
Secondly we discuss performance of the memory commenting on its tunability.

\subsection{Photon Generation}
\label{sec:photon_generation}
The HBT setup is used to characterize the photons generated by the source.
Photon arrival times at each SNSPD are recorded together with trigger times for each experimental trial.
We compute the second order autocorrelation function as:
\begin{equation}
    g^{(2)}(n)=\frac{c_{1,2}(n)}{p_{1}p_{2}},
\end{equation}
where $p_1$ ($p_2$) is the probability of detection per trial with SNSPD 1 (2) and $c_{1,2}(n)$ is the probability of a coincidence between detections separated by $n$ trials ($n=0$ means that detections are taken within the same trial).
All the probabilities are calculated within a detection time window at fixed delay after each trial trigger.
We choose a \SI{300}{\nano\second} detection window which includes more than \SI{95}{\percent} of the photon.
For perfect single photons $g^{(2)}(0) = 0$. In practice, background noise or multi-photon components increase the $g^{(2)}(0)$. Emitted light remains non classical for $g^{(2)}(0)<1$ and $g^{(2)}(0)=0.5$ marks the limit between single and multi-photon states. 

\begin{figure}[t]
	\includegraphics[width = \columnwidth]{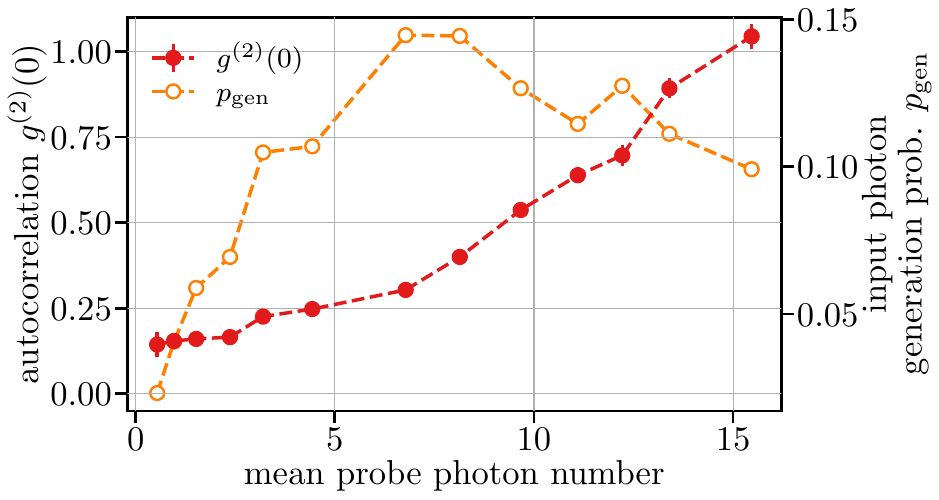}
	\caption{Dependence of $g^{(2)}(0)$ and $p_{\mathrm{gen}}$ on mean probe photon number.
	$p_{\mathrm{gen}}$ follows the Rabi cycle decreasing for the largest values of the probe power while $g^{(2)}(0)$ grows monotonically up to 1.}
	\label{fig:g2_rydberg}
\end{figure} 

In our source we can change the emitted photon $g^{(2)}(0)$ within a range of \numrange{0.16}{1} by varying the  mean probe photon number, see \autoref{fig:g2_rydberg}.
For smaller probe photon number values the increase of $g^{(2)}(0)$ is accompanied by an increase of the photon generation probability $p_{\mathrm{gen}}$, which is defined as $p_{\mathrm{gen}} = (p_1 + p_2) / \alpha$ where $\alpha = 0.21$ is the combined transmission and detection efficiency (of the source only - in this characterisation we detected photons right after the source).
However, for larger probe photon number $p_{\mathrm{gen}}$ decreases in accordance with the Rabi cycle.
Yet, this is not accompanied by the $g^{(2)}(0)$ which continues to grow up to 1, indicating the presence of multiphoton components.
The reason for this behaviour is still under investigation and goes beyond the scope of this paper.

If not stated otherwise, for the following measurements we fix $g^{(2)}(0) \approx 0.23$ and $p_{\mathrm{gen}} \approx \SI{12}{\percent}$.
The emitted photon has a steep leading edge followed by a slower exponential decay, with a FWHM of the entire photon of $\sim \SI{120}{\nano\second}$, see \autoref{fig:photons_and_storagetime} at time zero.

\subsection{Photon Storage}
\label{sec:photon_storage}
In this section we demonstrate that the memory can efficiently store the generated single photon. It also offers tunability in the storage and retrieval process.

\begin{figure}[!t]
	\includegraphics[width=\columnwidth]{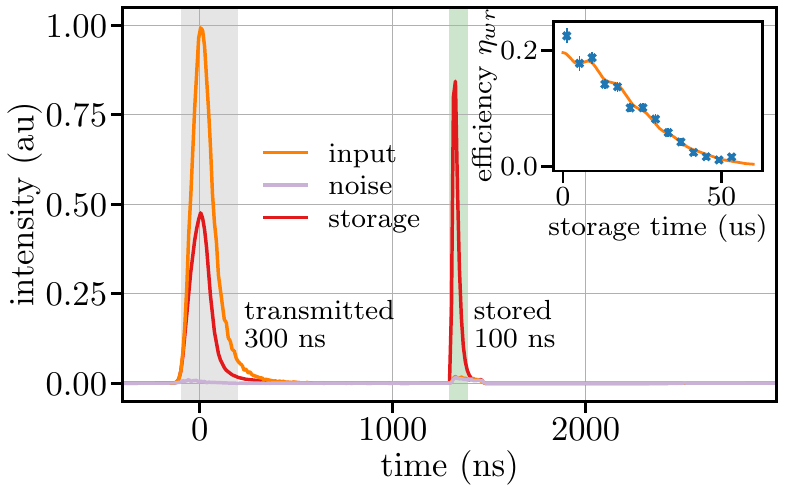}
	\caption{Photon histogram observed at the SNSPDs after the memory.
	The orange histogram is the input photon alone with no storage attempt.
	The red histogram presents a storage attempt.
	The lavender histogram shows the noise without input photon, but with the atoms in the memory.
	The grey shaded area is the detection window for the input and transmitted photon and green shaded area is the detection window for the stored photon.
	The inset shows the storage efficiency as a function of storage time with the corresponding Gaussian fit $e^{-t^2/\tau^2}$, where $\tau$ is the memory lifetime.
	The fit also includes an oscillatory term accounting for spin wave interference coming from residual population in $\ket{5S_{1/2}, F=2, m_F=1}$ (as an effect of imperfect Zeeman optical pumping) \cite{Albrecht2015}.
	}
	\label{fig:photons_and_storagetime}
\end{figure}

To characterize the memory performance, we first measure the temporal histogram of photon counts in 3 different situations, as shown in \autoref{fig:photons_and_storagetime}.
We first detect the input single photon (orange histogram) when no storage attempt is performed, i.e. with no atoms in the memory but with control pulses.
Then, a storage attempt is performed (red histogram) and one can see two peaks - the transmitted photon, which is the part of the input photon that is not absorbed in the storage attempt (counts in the \SI{300}{\nano\second} grey shaded window), and the stored photon, which is the excitation retrieved from a successful storage attempt (counts in the \SI{100}{\nano\second} green shaded window). 
Finally, we measure the noise (lavender histogram) by blocking the input photon while keeping all the control pulses on and the atomic cloud present.
This last measurement should contain all the information about the noise present in the experiment, in particular the noise introduced by the control pulses.
We measure a noise probability per trial within the storage window of $p_n= 2.3(3) \times 10^{-4}$, which, to our knowledge, is among the lowest reported in ground state spin-wave memories.

The input and the noise histograms serve as a reference to calculate the storage efficiency $ \eta_{wr} = p_{\mathrm{s}} / p_{\mathrm{in}}$, where $p_{\mathrm{s}}$ and $p_{\mathrm{in}}$ are background-subtracted probabilities of detecting a stored photon (within the \SI{100}{\nano\second} detection window) and an input photon (within the \SI{300}{\nano\second} detection window), respectively.
We also calculate the write-in efficiency defined as $\eta_w =  (p_{\mathrm{in}} - p_{\mathrm{t}}) / p_{\mathrm{in}}$, where $p_{\mathrm{t}}$ is the background-subtracted detection probability of a transmitted photon (within the \SI{300}{\nano\second} detection window).
From these two quantities we infer the read-out efficiency $\eta_r = \eta_{wr} / \eta_w$.
We obtain a maximum storage efficiency $\eta_{wr} \approx \SI{21}{\percent}$ at a storage time of \SI{1.2}{\micro\second}.
For longer storage times the motional decoherence and the decoherence due to the stray magnetic field gradients limits the efficiency with a characteristic $1/e$ decay time of \SI{30}{\micro\second} (see inset in \autoref{fig:photons_and_storagetime}).

For the measurement shown in \autoref{fig:photons_and_storagetime}, the SNR of the retrieved photon is 24(4). For different input number of photons, we measure SNR of up to 26 (see supplemental materials \cite{SupMat}).
An interesting figure of merit is the $\mu_1$ parameter, defined as  $\mu_1=p_n / \eta_{wr}$, which expresses the input number of photons required to have $\text{SNR}=1$ at the output.
In our case, we find  $\mu_1 = \num{1.00(7)e-3}$ (see supplemental materials \cite{SupMat}), which is more than two orders of magnitude lower than similar ground state quantum memories based on warm atomic vapors \cite{Wolters2017, Thomas2019, Namazi2017}, and more than one order of magnitude lower than solid-state QMs based on rare-earth doped solids \cite{Seri2017, Gundogan2015}. 

\begin{figure}[!t]
	\includegraphics[width=\columnwidth]{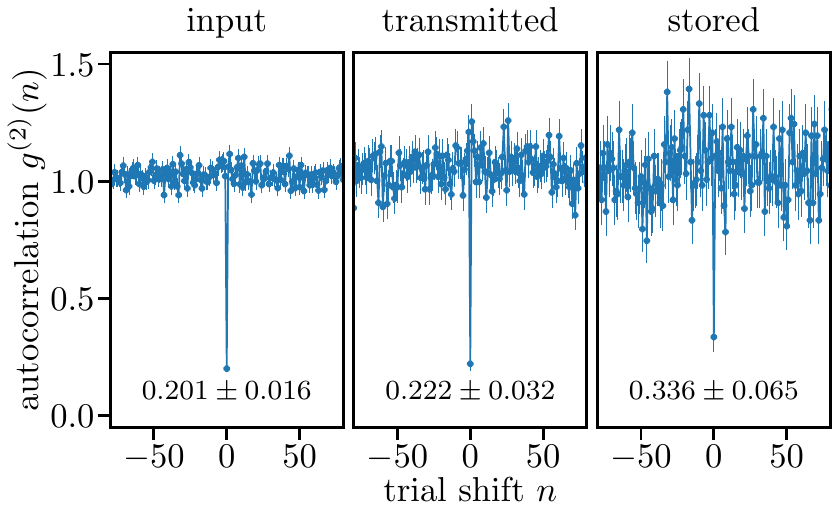}
	\caption{Autocorrelation $g^{(2)}$ as a function of shift between trials $n$ for the input, transmitted and stored photons.
	For trials separated by a shift $n \geq 1$ the clicks are uncorrelated yielding $g^{(2)}(n) = 1$.
	Coincidences clicks in the same trial, $n = 0$, are much less frequent asserting the photons anti-bunching.}
	\label{fig:g2}
\end{figure}

A crucial requirement for a quantum memory is that it preserves the statistical properties of the stored photons.
To show that our memory fulfills this criterion, we first adjust the mean probe photon number of the source to low values, resulting in a measured photon generation probability of $p_{\mathrm{gen}} \approx \SI{3.0(3)}{\percent}$ (see \autoref{fig:g2_rydberg}).
With this setting, we expect the emitted photons to be strongly non-classical.
To reduce the effect of experimental fluctuations, we collect data for 63 hours.
We measure $g^{(2)}(0)$ of the input ($g^{(2)}(0)=0.20(2)$), transmitted  (($g^{(2)}(0)=0.22(3)$) and stored photons ($g^{(2)}(0)=0.34(7)$) and obtain values well below \num{0.5}, see \autoref{fig:g2}.
It shows that the memory preserves the single photon nature of the input photon.
One can see, however, that $g^{(2)}(0)$ of the stored photon is significantly larger than  the $g^{(2)}(0)$ of the input photon.
We expect that the main source of degradation of $g^{(2)}(0)$ is the uncorrelated noise introduced by the memory control pulses.
We developed a simple model, discussed in the supplemental materials \cite{SupMat}, to quantify the effect of uncorrelated noise on $g^{(2)}(0)$.
The model predicts a $g^{(2)}_m(0)=0.33 (4)$ for stored photon taking into account a measured SNR of $11(2)$ and the measured input $g^{(2)}(0)$.
For this data set, the model is in agreement with the measured data, within the error bars.
We also performed several other measurements (see  supplemental materials \cite{SupMat}, with integration times of around 10-16 hours per data point) for different input $g^{(2)}(0)$.
While the model reproduces qualitatively the trend, there is a large point to point fluctuation that we attribute to low statistics and experimental fluctuations. 

\begin{figure}[!b]
	\includegraphics[width=\columnwidth]{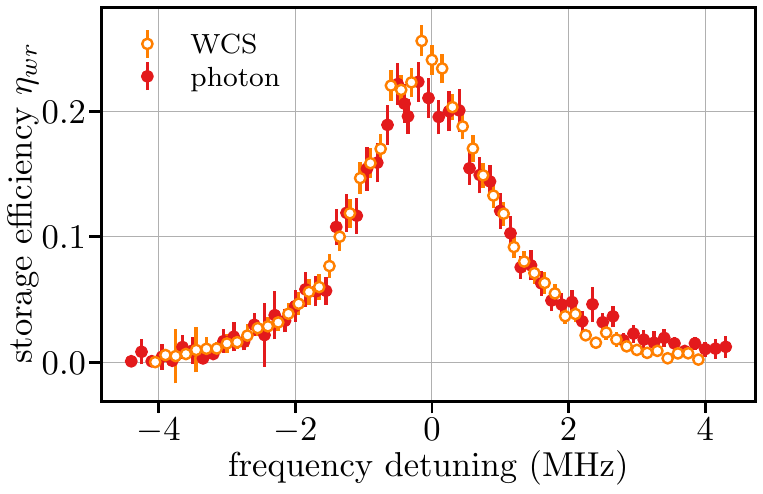}
	\caption{Storage efficiency versus frequency detuning of the control write-in pulse for the single photon input (red) and WCS (orange).
	The frequency detuning is measured from the two-photon resonance of the input photon and the write-in control beam.}
	\label{fig:spectrum}
\end{figure}

Our memory offers significant tunability in the write-in process that may prove useful in future hybrid quantum networks \cite{Maring2017}.
We start by showing that the memory can adapt to the input photon frequency.
For that we vary the frequency of the control beam pulse.
The maximum efficiency is observed for the two-photon resonance, see \autoref{fig:spectrum}, achieving optimum storage conditions for the input photon.
The width of the curve depends on the spectral properties of the input photon.
Bandwidth-limited photons (i.e. photons that exhibit the minimum bandwidth for a given temporal duration) are desirable because one can achieve with them a high Hong-Ou-Mandel interference visibility \cite{Hong1987} which is a requirement for most of the entanglement distribution protocols. 
Therefore, in order to  benchmark the spectral properties of the input photon, we repeat the measurement with a weak coherent state (WCS) with the same waveshape, center frequency and mean number of photons. 
This WCS is derived from a laser exhibiting a linewidth much smaller than the bandwidth of the pulse.
As can be seen in \autoref{fig:spectrum}, both spectra overlap very well suggesting that the input photon is close to bandwidth-limited.

We also study the bandwidth of our memory by sending WCS of different durations (with FWHM of \SIrange{8}{800}{\nano\second}).
We show the results in the supplemental materials \cite{SupMat} asserting that the memory can accommodate pulses of very different lengths without changing its efficiency.

\begin{figure}[!t]
	\includegraphics[width=\columnwidth]{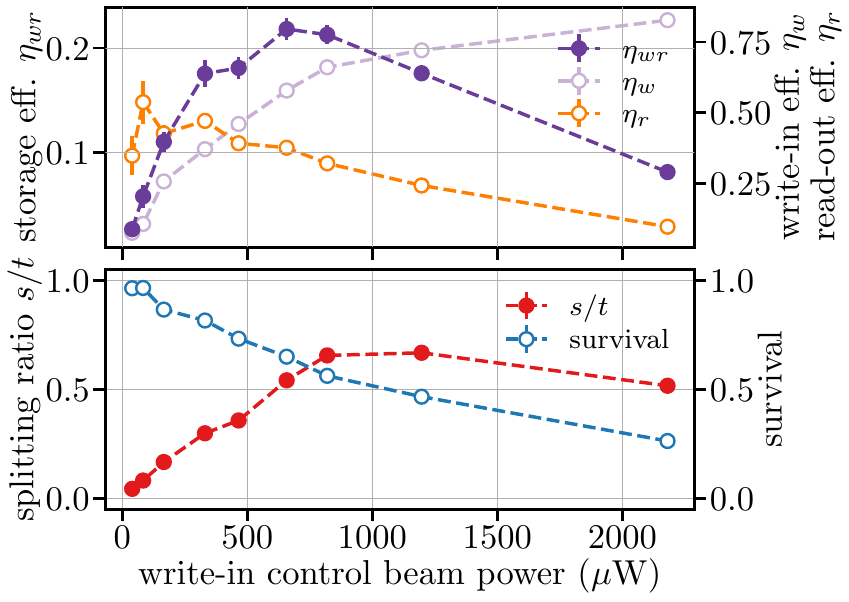}
	\caption{Storage efficiency $\eta_{wr}$ and related quantities as a function of the write-in control beam power. In the top panel we show the efficiencies $\eta_{w}$, $\eta_{r}$ and $\eta_{wr}$ and in the bottom panel we show the survival efficiency and $s/t$.}
	\label{fig:survival}
\end{figure}

Another interesting feature of the memory is that one can control how much of the input photon is absorbed and how much is transmitted.
By varying the write-in control beam power one can change $\eta_w$ as shown in \autoref{fig:survival} (top).
This changes effectively the splitting ratio $s/t$ between the stored and the transmitted photon pulse
\begin{equation}
    \frac{s}{t} \defeq \frac{\eta_{wr}}{1 - \eta_w} = \frac{p_{\mathrm{s}}}{p_{\mathrm{t}}}.
\end{equation}
Our memory can therefore be used as a temporal beam splitter\cite{Reim2012} with a tunable splitting ratio, which may have applications in the quantum repeater architecture mentioned in the introduction. To investigate this possibility, we plot $s/t$, see \autoref{fig:survival} (bottom).
We see that $s/t$ peaks for intermediate values of the write-in control beam powers and decays for higher values.
This stands in contrast with the monotonically growing $\eta_w$ and is a result of $\eta_r$ decreasing with the control power.
We attribute this behaviour of $\eta_r$ to the asymmetrical distribution of the spin wave in the ensemble, when large write-in control powers are used \cite{Nunn2007,Gorshkov2007a}.
With increasing write power, the spin wave starts having more asymmetric shape, being mostly created at the beginning of the ensemble. 
This effect is known to limit the retrieval efficiency, especially in the forward retrieval configuration \cite{Nunn2007, Vernaz-Gris2018a}. 

In \autoref{fig:survival} (bottom), we also plot the survival efficiency, defined as the probability of detecting a transmitted or stored photon per trial.
We observe that it decreases with increasing control power due to the decrease of the read-out efficiency.
With current conditions, the tunability range of $s/t$ is limited, but we expect that backward retrieval should considerably improve the read-out efficiency at high write power, which will increase the survival probability \cite{Gorshkov2007a}.
As a first application of the single photon temporal beam splitter, we used the two temporal output modes of the memory to measure the antibunching parameter.
For the measurement presented in \autoref{fig:g2}, we obtain a $g^{(2)}(0)= 0.28(2)$ with a significantly increased count rate with respect to the case where we split each output mode with a standard BS.

\begin{figure}[t]
	\includegraphics[width=0.98\linewidth]{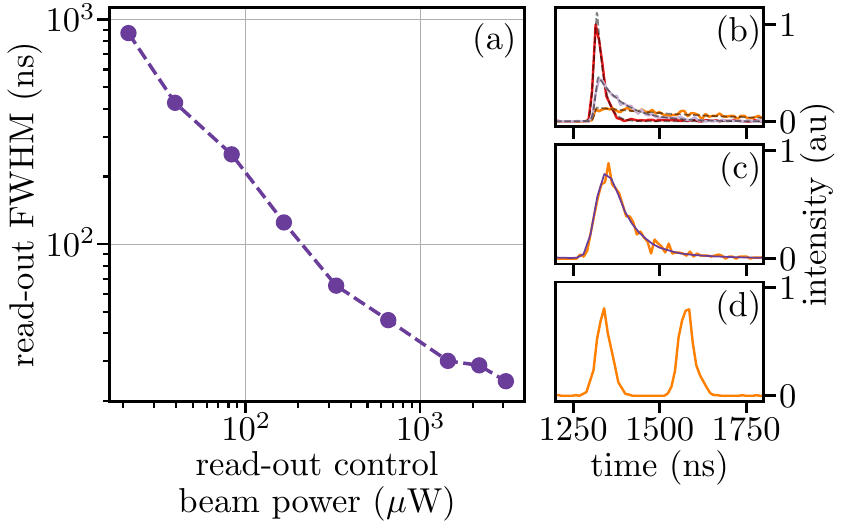}
	\caption{Stored photon waveshape tunability.
	(a) Dependence of the stored photon duration on the read-out control beam power. 
	(b) Selected waveshapes of the stored photons from (a) and their corresponding fits.
	(c) Stored photon waveshape (solid orange) matching exactly the input photon waveshape (dashed blue).
	(d) Stored photon shaped as a time-bin qubit.
	}
	\label{fig:different_readout_durations}
\end{figure}

Our memory also offers shape tunability of the stored photon \cite{Farrera2016a}.
In particular, one can retrieve photons with very different lengths (with FWHM of \SIrange{25}{900}{\nano\second}) by changing the read-out control beam power, see \autoref{fig:different_readout_durations}(a, b).
We read out the memory with a square-shaped pulse resulting in a steep leading edge of the retrieved photon and slower decaying trailing edge.
We fit the former with a Gaussian function and the latter with an exponential and obtain the total FWHM of the photon.
One can also use more complex waveforms for the read-out control pulse to shape the read-out photon, e.g. reproducing the input photon or a time-bin qubit, see \autoref{fig:different_readout_durations}(c, d).
This capability would allow for matching differently shaped photons emitted by different sources.
We do not observe significant reduction of $\eta_{wr}$ for different read-out pulse shapes, in agreement with theory \cite{Nunn2008a}.

\section{Conclusion}
\label{sec:conclusions}

We demonstrated storage and retrieval of an on-demand single photon generated in one Rydberg-based atomic ensemble in another cold atomic ensemble trough a Raman memory protocol.
We achieved a \SI{21}{\percent} memory efficiency and a signal-to-noise ratio up to 26 for the retrieved photon, leading to $\mu_1$ of \num{1.00(7)e-3}.
This allowed us to observe only a moderate degradation of the single photon statistics.
We showed the adaptability of our memory in frequency and bandwidth.
Moreover, we evaluated the performances of the built-in temporal beam splitter offered by the Raman memory. 
Lastly, we showed that we can shape the temporal waveform of the retrieved  photon by shaping the read-out pulse power and waveform. 
These results are a step forward in the implementation of efficient quantum-repeater links using single-photon sources.
In that context, one interesting advantage of having the source and the memory residing in different ensembles is that they can be optimized independently.
This allows for an efficient single-photon generation and storage and facilitates the use of multiplexed quantum memories \cite{Heller2020, Pu2017}, which would significantly improve repeater entanglement generation rates. 

Several improvements should be applied to our experiment before it can become a practical alternative.
The generation efficiency of the single photon from the Rydberg ensemble could be increased by increasing the OD of the ensemble and/or by embedding the ensemble in a low finesse cavity \cite{Yang2021}.
The quality of the single photon (as measured by the autocorrelation function $g^{(2)}(0)$) could also be improved by addressing a slightly smaller ensemble and by reaching higher principal quantum number level to increase the Rydberg blockade radius, as was shown in \cite{OrnelasHuerta2020}, where $g^{(2)}(0)$ values smaller than $10^{-3}$ have been measured.
Regarding the Raman quantum memory, higher storage and retrieval efficiencies could also be reached by increasing the OD of the ensemble \cite{Cao2020} and using backward retrieval \cite{Vernaz-Gris2018a}, or with an impedance matched cavity.
Backward retrieval will also improve the survival probability and the performances of the temporal beam-splitter.
Finally, longer storage time up to 1 s could be achieved by using magnetic insensitive transitions and by loading the ensemble into an optical lattice to suppress motional induced dephasing \cite{Wang2021}. 

\emph{Note added}. While finalizing our experiment, we learned about a recent experiment where a single photon generated by Rydberg atoms was stored in an atomic ensemble using electromagnetically induced transparency \cite{Yu2021}.

\begin{acknowledgements}

We acknowledge interesting discussions with Gerhard Rempe.

L.H. and J.L. acknowledge funding from the European Union’s Horizon 2020 research and innovation programme under the Marie Sk\l{}odowska-Curie grant agreement No. 713729.

This project received funding from the Government of Spain (PID2019-106850RB-I00 project funded by MCIN/ AEI /10.13039/501100011033; Severo Ochoa CEX2019-000910-S), from the European Union's Horizon 2020 research and innovation program under Grant Agreement No. 899275 (DAALI), from the Gordon and Betty Moore Foundation through Grant No. GBMF7446 to H. d. R, from Fundaci{\'o} Cellex, Fundaci{\'o}  Mir-Puig and from   Generalitat de Catalunya (CERCA, AGAUR)

L.H. and J.L. contributed equally to this work.

\end{acknowledgements}

%

\end{document}